\documentclass[submission,copyright,creativecommons]{eptcs}
\providecommand{\event}{QSQW 2020}
\usepackage{underscore}
\usepackage{graphicx}
\usepackage{amsmath,amssymb}
\usepackage{cite}

\def \C {\mathbb C}

\def \R {\mathbb R}

\def \Z{\mathbb Z}

\newtheorem{definition}{Definition}[section]
\newtheorem{proposition}{Proposition}[section]

\makeatother
\makeatletter
\title{Bifurcation Curves of Two-Dimensional Quantum Walks}
\author{
Parker Kuklinski
\institute{Naval Undersea Warfare Center, USA}
\and
Mark Kon
\institute{Department of Mathematics and Statistics, Boston University, Boston, USA}
\institute{Department of Mathematics, Massachusetts Institute of Technology, Cambridge, MA}
}
\def\titlerunning{Bifurcation Curves of Two-Dimensional Quantum Walks}
\def\authorrunning{P. Kuklinski \& M. Kon}

\begin{document}
\maketitle

\begin{abstract}
The quantum walk differs fundamentally from the classical random walk in a number of ways, including its linear spreading and initial condition dependent asymmetries. Using stationary phase approximations, precise asymptotics have been derived for one-dimensional two-state quantum walks, one-dimensional three-state Grover walks, and two-dimensional four-state Grover walks. Other papers have investigated asymptotic behavior of a much larger set of two-dimensional quantum walks and it has been shown that in special cases the regions of polynomial decay can be parameterized. In this paper, we show that these regions of polynomial decay are bounded by algebraic curves which can be explicitly computed. We give examples of these bifurcation curves for a number of two-dimensional quantum walks.
\end{abstract}

\section{Introduction}
The quantum walk is a discrete quantum mechanical system which serves as an analogue to the classical random walk. While the probability distribution of a classical random walk approximates a normal distribution with a linearly growing variance in large time, the quantum walk has more involved asymptotic behavior. A paper by Ambainis et. al. \cite{ambainis01} provides a detailed asymptotic description of the one-dimensional Hadamard walk (a quantum walk governed by a $2\times 2$ Hadamard matrix) for large time. The large time behavior of the quantum walk contrasts with that of the classical random walk in both the shape of the probability density and its rate of spread. Ambainis et. al. showed that the standard deviation of the Hadamard walk grows at $O(t)$ and in particular the distribution at time $t$ is almost entirely contained in the interval $\left[-\frac{t}{\sqrt{2}},\frac{t}{\sqrt{2}}\right]$. In this interval, the distribution contains highly oscillatory peaks at the boundaries and is approximately uniform near the origin.

These results were among the first in a line of papers describing asymptotic behavior of more general quantum walks. A few years later, Konno et. al. \cite{konno02} extended these results to two state quantum walks governed by general unitary matrices. Inui et. al. \cite{inui05} provided asymptotic results on the three-state one dimensional Grover walk with a focus on localization phenomena. Analysis of two-dimensional quantum walks started with a brief survey by Mackay et. al. \cite{mackay02}, and continued with a description of localization in the two-dimensional Grover walk by Inui et. al. \cite{inui04}, and a differential geometric interpretation of two-dimensional quantum walks by Baryshnikov et. al. \cite{baryshnikov11}. The asymptotic behavior of the one dimensional Hadamard walk described by Ambainis et. al. has served as a common baseline through these papers. The standard deviation of these walks grows as $O(t)$, and the probability distributions are almost entirely contained in a linearly expanding subset of the domain, which we term as the region of polynomial decay. In Kuklinski \cite{kuklinski17}, these regions of polynomial decay are investigated further. In particular, it was shown that in certain low state two-dimensional quantum walks, the regions of polynomial decay can be explicitly parameterized. However, these parameterizations are often unwieldy and cannot easily give a description of the bounds on the region.

In this paper, we show that for any quantum walk defined on $\Z ^d$, one can write down a collection of algebraic curves which bound the regions of polynomial decay. This procedure builds upon previous methods to analyze the asymptotic behavior of the quantum walk. First, we conduct an eigenvalue decomposition of the quantum walk operator via a Fourier transform. The state of the quantum walk particle at time $t$ becomes a sum of integrals whose asymptotic behavior we analyze using the method of stationary phase. Solving for points of stationary phase in these integrals leads to the previously discussed parametric representation of the regions of polynomial decay. We will take this a step further and compute bifurcation curves corresponding to these integrals by computing a Hessian determinant of the phase of the integrand. These bifurcation curves divide the space $\R^d$ into a discrete collection of subsets, a finite number of which are found to belong to the region of polynomial decay. The bifurcation curves are found to be solutions to a system of multivariate polynomial equations. We use a Gr{\"o}bner basis computation to derive an implicit algebraic representation of these curves. However, this system of multivariate polynomial equations is quite large which puts strain on our algorithm, thus only in the simplest cases can we derive the bifurcation curves. We present a non-rigorous ad-hoc method for computing bifurcation curves for more complicated quantum walks.

The remainder of this paper is organized as follows. The definition of the quantum walk as well as a discussion of stationary phase approximations is given in section \ref{sec:2}. In section \ref{sec:3} we explicitly write the multivariate polynomial system corresponding to the bifurcation curves and discuss solution techniques. In section \ref{sec:4} we compute bifurcation curves for several examples of two-dimensional quantum walks.

\section{Definitions and Method of Stationary Phase} \label{sec:2}
We begin by defining the quantum walk on a group, as first introduced by Acevedo et. al. \cite{acevedo05}:
\begin{definition}
Let $(G,\cdot )$ be a group, let $\Sigma\subset G$ with $|\Sigma |=n$, and let $U$ be an $n\times n$ unitary matrix. The quantum walk operator $Q:\ell ^2(G\times\Sigma )\rightarrow\ell ^2(G\times\Sigma )$ corresponding to the triple $(G,\Sigma ,U)$ may be written as the composition $Q=T(I\otimes U)$ where for $g\in G$ and $\sigma\in\Sigma$, $T:|g\rangle |\sigma\rangle\mapsto |\sigma\cdot g\rangle |\sigma\rangle$. We denote this correspondence as $Q\leftrightarrow (G,\Sigma ,U)$.
\end{definition}
The ordered pair $(G,\Sigma )$ can be thought of as an undirected Cayley graph which admits loops \cite{diestel05}. In this paper, we will primarily consider $G=\Z ^2$. Let $C_d\subset\Z^d$ be the set of unit directional vectors and let  $\tilde{C_d}=C_d\cup \{ 0\}$. Two of the unitary matrices which we will use in this paper are the Grover matrix and the Hadamard matrix. If $I_n$ is the $n\times n$ identity matrix and $\text{\bf 1}_n$ is the $n\times n$ matrix filled with ones, then we define the $n\times n$ Grover matrix as $G_n=\frac{2}{n}\text{\bf 1}_n-I_n$. The Hadamard matrix is defined as $H=\frac{1}{\sqrt{2}}\begin{bmatrix} 1 & 1 \\ 1 & -1\end{bmatrix}$.

One way to view this quantum walk operator is as a linear combination of translations. Let $\psi\in\ell ^2(G)$ and let $T_\sigma :\ell ^2(G)\rightarrow\ell ^2(G)$ be a translation operator which acts as $T_\sigma (\psi (g))=\psi (\sigma ^{-1}\cdot g)$ with $\sigma\in\Sigma$. Then we can visualize $Q$ acting on the vector $\Psi =\left[ \psi _{\sigma _1},...,\psi _{\sigma _n}\right]$ as follows:
\begin{equation}
Q\Psi =\begin{bmatrix} T_{\sigma _1} & ~ & ~ \\ ~ & \ddots & ~ \\ ~ & ~ & T_{\sigma _n}\end{bmatrix}U\Psi
\end{equation}

When $G=\Z^d$, we can gain a better understanding of $Q$ through the application of a discrete Fourier transform. If $\theta\in\R^d$, the multi-dimensional Fourier transform acts on a translation as:
\begin{equation}
\mathcal{F}[T_\sigma\psi ](\theta )=e^{i\sigma\cdot\theta}\mathcal{F}[\psi ](\theta )
\end{equation}
Applying a Fourier transform to equation (1) and using equation (2), we find:
\begin{equation}
\mathcal{F}[Q\Psi ](\theta )=\begin{bmatrix} e^{i\sigma _1\cdot\theta} & ~ & ~ \\ ~ & \ddots & ~ \\ ~ & ~ & e^{i\sigma _n\cdot\theta}\end{bmatrix}U\mathcal{F}[\Psi ]=M(\theta )\mathcal{F}[\Psi ](\theta )
\end{equation}
We refer to $M(\theta )$ as the \emph{multiplier matrix} of the quantum walk operator.

After $t$ steps of the quantum walk, the Fourier transform of the state becomes \\
$\mathcal{F}[Q^t\Psi ](\theta )=M(\theta )^t\mathcal{F}[\Psi ](\theta )$. Thus, if we wish to study long term behavior of the quantum walk, we must take large powers of the multiplier matrix. If $\{ \lambda _m (\theta )\}_{m=1}^n$ is the set of eigenvalues of $M(\theta )$, then we can write the transform of a quantum walk with initial condition $\Psi _0$ as:
\begin{equation}
M(\theta )^t\mathcal{F}[\Psi _0](\theta )=\sum _{m=1}^n\lambda _j(\theta )^tE_m(\theta )
\end{equation}
Here, the $E_m (\theta )$ are scaled eigenvectors of $M(\theta )$ whose weights depend on the initial condition. Since $M(\theta )$ is unitary, we can write $\lambda _m (\theta )=e^{iH_m (\theta )}$ where $H_m:[-\pi ,\pi ]^n\rightarrow\R$. To return to the original domain, we conduct an inverse Fourier transform on equation (4):
\begin{equation}
Q^t\Psi _0=\frac{1}{(2\pi )^d}\sum _{m=1}^n\int _{\lVert\theta\rVert _\infty\le\pi}E_m(\theta )\exp\left[ i(tH_m(\theta )-x\cdot\theta )\right] d\theta
\end{equation}
Here, $\lVert\cdot\rVert _\infty$ refers to the $\ell ^\infty$ norm \cite{kreyszig78}. We let $x=Xt$ such that our spatial variable of interest is now $X\in\R^d$. This scales position space such that we will no longer be observing a linearly expanding spatial region but a stationary one. Substituting this into equation (5), we have:
\begin{equation}
Q^t\Psi _0=\frac{1}{(2\pi )^d}\sum _{m=1}^n\int _{\lVert\theta\rVert _\infty\le\pi}E_m(\theta )\exp\left[ it(H_m(\theta )-X\cdot\theta )\right] d\theta
\end{equation}

We use the method of stationary phase \cite{bleistein75} to asymtptotically evaluate the integrals in equation (6). Consider the following $d$-dimensional oscillatory integral:
\begin{equation}
I(t)=\int _{\R ^d}g(x)e^{itf(x)}dx .
\end{equation}
where $g$ and $f$ are smooth functions with compact support. Consider the set:
$$S=\left\{ x\in\R^d:\nabla f(x)=0,\det\left(\left[\frac{\partial ^2f}{\partial x_j\partial x_k}\right]\right)\ne 0\right\}$$
whose members we refer to as \emph{points of stationary phase} or \emph{nondegenerate critical points}. If $\nabla f(x)=0$ and the determinant of the Hessian matrix also vanishes, we say that $x$ is a \emph{degenerate critical point}. We state two results from Stein \cite{stein93} relating this set to an approximation of $I(t)$ in equation (7):
\begin{proposition}
Suppose $f$ has no critical points in the support of $g$. Then $I(t)=O(t^{-n})$ for every $n\ge 0$.
\end{proposition}
This proposition says that if the set of critical points is empty, then the integral $I(t)$ in equation (7) decays superpolynomially (often this superpolynomial decay can be shown to be exponential). If the set of critical points is nonempty, then we can use the following proposition:
\begin{proposition}
If $f$ has a nondegenerate critical point at $x_0$ and $g$ is supported in a sufficiently small neighborhood of $x_0$, then $I(t)=O(t^{-d/2} )$.
\end{proposition}
Thus, if the set of nondegenerate critical points is nonempty, then the integral $I(t)$ decays polynomially, slower than the decay dictated by proposition 1.

We apply these propositions to the integrals in equation (6). Let $f_j (\theta )=H_j (\theta )-X\cdot\theta$ correspond to the expression in the exponential for the $j^\text{th}$ integrand. This integral has a critical point if $\nabla f_j (\theta )=0$, or if $\nabla H_j (\theta )=X$ for some $\theta \in [-\pi ,\pi]^d$. By letting $\theta$ vary in this range we can trace out a region in $\R^d$ on which the amplitudes decay polynomially. In this way, $\nabla H_j$ maps the region of integration $[-\pi ,\pi]^d$ into the spatial domain on which the quantum walk resides. This representation requires that we can analytically solve for the eigenvalues $\lambda _m (\theta )$ of $M(\theta )$ which is not always possible, especially for quantum walks with many states. Moreover, this parametric representation is often complicated and it is not immediately apparent how to connect this representation to the more general structure of the region of polynomial decay, or even to a mathematical description of its boundaries.

We mention that a critical condition for the stationary phase propositions is the smoothness imposed on $f$ and $g$. In the context of proposition 2.1, a discontinuity in an $n^\text{th}$ derivative of one of these functions will lead to a slower polynomial rate of decay than the superpolynomial decay guaranteed by the proposition for smooth amplitude and phase functions. It is not trivial to show that the eigenvalues and eigenvectors of the multiplier matrix are smooth. The following proposition was proved by Rainer \cite{rainer13}:
\begin{proposition}
Let $A(t) = [A_{ij} (t)]_{1\le i,j\le n}$ be a $\mathcal{C}$-curve of normal complex matrices, i.e., the entries $A_{ij}$ belong to $\mathcal{C}(\R, \C)$, such that $P_A$ is normally nonflat. Then there exists a global $\mathcal{C}$-parameterization of the eigenvalues and the eigenprojections of $A$.
\end{proposition}
Here, $\mathcal{C}$ is a subalgebra of $C^\infty$ (Rainer notes that it may be true that $\mathcal{C}=C^\infty$), which the entries of $M(\theta )$ can be shown to belong to. However, this proposition only applies to families of normal matrices whose entries are smothly parameterized by a single variable. Families of normal matrices parameterized by more than one variable may not admit a smooth selection of eigenvalues (e.x. consider the family of Hermitian matrices $\begin{bmatrix} x & y \\ y & -x\end{bmatrix}$). This means that for one-dimensional quantum walks, we can provably distinguish between a region of polynomial decay and a region of superpolynomial decay. This distinction is not guaranteed for higher dimensional quantum walks, although in the examples that follow a smooth selection of eigenvalues and eigenvectors can be demonstrated. In any case, for general higher dimensional quantum walks a lack of smoothness in the eigenvalues and eigenvectors of the multiplier matrix would not negate existence of a distinct region of polynomial decay bounded by algebraic curves, it simply does not guarantee qualitatively faster rates of decay outside this region.

\section{Multivariate Polynomial System} \label{sec:3}
Instead of focusing our attention on the locus of critical points, we illustrate a method to construct algebraic surfaces representing the bifurcation curves of the quantum walk. These bifurcation curves will more appropriately describe the structure of the region of polynomial decay than does the aforementioned critical point parameterization. The degenerate critical points $\theta$ of the integral in equation (6) must simultaneously satisfy:
\begin{equation}
\det\left(\left[\frac{\partial ^2 H_m}{\partial \theta_j\partial \theta_k}\right] (\theta )\right) =0
\end{equation}
\begin{equation}
\nabla H_m(\theta )=X
\end{equation}

Using implicit differentiation, one can represent equations (8) and (9) as multivariate polynomial equations in terms of derivatives of the characteristic polynomial. Let $p_0 (\lambda ;\theta _k )$ be the characteristic polynomial of $M(\theta )$ in $\lambda$. Since the entries of $M(\theta )$ are linear combinations of terms of the form $e^{i\theta _k}$, we can write a multivariate polynomial $p(\lambda ;x_k )$ such that the roots in $\lambda$ of $p(\lambda ;e^{i\theta _k } )$ coincide with the eigenvalues of $M(\theta )$ (i.e. the roots of $p_0(\lambda ;\theta _k)$ in $\lambda$). Thus, if $\lambda (\theta )$ is an eigenvalue of $M(\theta )$, we have:
\begin{equation}
p\left(\lambda (\theta );x_k(\theta )\right) =0
\end{equation}
Recall that $\lambda (\theta )=e^{iH(\theta )}$, and that we are searching for derivatives of $H(\theta )$ to use in equations (8) and (9). As such, let us take a derivative of this equation with respect to $\theta _j$:
\begin{equation}
\frac{\partial\lambda}{\partial\theta _j}=i\frac{\partial H}{\partial\theta _j}e^{iH(\theta )}=i\frac{\partial H}{\partial\theta _j}\lambda (\theta )
\end{equation}
From here, let us refer to partial derivatives via subscripts (not to be confused with the indexing of eigenvalues in the previous section) and suppress mention of $\theta$. Rearranging terms, we find:
\begin{equation}
H_j=\frac{\lambda _j}{i\lambda}
\end{equation}
If $X=(X_1,...,X_n )$, we can substitute equation (12) into equation (9) to find:
\begin{equation}
\lambda _j-i\lambda X_j=0
\end{equation}
If we take a second derivative of equation (11) with respect to $\theta _k$, we have:
\begin{equation}
\lambda _{jk}=i(H_{jk}\lambda +H_j\lambda _k)
\end{equation}
By isolating $H_{jk}$ and substituting the expression in equation (12) for $H_j$, we may write:
\begin{equation}
H_{jk}=\frac{i}{\lambda ^2}\left(\lambda _j\lambda _k-\lambda _{jk}\lambda\right)
\end{equation}
We can similarly substitute equation (15) into equation (8) such that:
\begin{equation}
\det\left(\left[\lambda _j\lambda _k-\lambda _{jk}\lambda\right]\right) =0
\end{equation}

Both equations (13) and (16) may be used to describe the bifurcation curves, however these equations are dependent on derivatives of $\lambda (\theta )$. We can solve for these derivatives in terms of the characteristic polynomial by taking derivatives of equation (10) with respect to $\theta _j$. Let us take a first derivative with respect to $\theta _j$:
\begin{equation}
\frac{\partial p}{\partial\lambda}\frac{\partial\lambda}{\partial\theta _j}+\frac{\partial p}{\partial x_j}\frac{\partial x_j}{\partial\theta _j}=0
\end{equation}
Noticing that $\frac{\partial x_j}{\partial\theta _j}=ix_j$, we can rearrange terms to write:
\begin{equation}
\lambda _j=-\frac{ip_jx_j}{p_\lambda}
\end{equation}
This can be substituted into equation (13) to find:
\begin{equation}
p_jx_j+\lambda p_\lambda X_j=0
\end{equation}
We take an additional derivative of equation (17) with respect to $\theta _k$. If $j\ne k$, then we can write:
\begin{equation}
\left(\frac{\partial p_\lambda}{\partial\lambda}\frac{\partial\lambda}{\partial\theta _k}+\frac{\partial p_\lambda}{\partial x_k}\frac{\partial x_k}{\partial\theta _k}\right)\frac{\partial\lambda}{\partial\theta _j}+\frac{\partial p}{\partial\lambda}\frac{\partial ^2\lambda}{\partial\theta _j\partial\theta _k} +\left(\frac{\partial p_j}{\partial\lambda}\frac{\partial\lambda}{\partial\theta _k}+\frac{\partial p_j}{\partial x_k}\frac{\partial x_k}{\partial\theta _k}\right)\frac{\partial x_j}{\partial\theta _j}=0
\end{equation}
If we rearrange terms and substitute equation (18), we have:
\begin{equation}
\lambda _{jk}=\frac{x_jx_k}{p_\lambda ^3}\left[ p_{\lambda\lambda}p_jp_k-p_{\lambda k}p_jp_\lambda -p_{\lambda _j}p_kp_\lambda +p_{jk}p_\lambda ^2\right]
\end{equation}
Meanwhile, if $j=k$ then we must account for an additional term:
\begin{equation}
\left(\frac{\partial p_\lambda}{\partial\lambda}\frac{\partial\lambda}{\partial\theta _j}+\frac{\partial p_\lambda}{\partial x_j}\frac{\partial x_j}{\partial\theta _j}\right)\frac{\partial\lambda}{\partial\theta _j}+\frac{\partial p}{\partial\lambda}\frac{\partial ^2\lambda}{\partial\theta _j^2} +\left(\frac{\partial p_j}{\partial\lambda}\frac{\partial\lambda}{\partial\theta _j}+\frac{\partial p_j}{\partial x_j}\frac{\partial x_j}{\partial\theta _j}\right)\frac{\partial x_j}{\partial\theta _j}+\frac{\partial p}{\partial x_j}\frac{\partial ^2x_j}{\partial\theta _j^2}=0
\end{equation}
Expanding this expression and making similar substitutions, we find:
\begin{equation}
\lambda _{jj}=\frac{x_j}{p_\lambda ^3}\left[ p_{\lambda\lambda}p_j^2x_j-2p_{\lambda j}p_jp_\lambda x_j+p_{jj}p_\lambda ^2x_j+p_jp_\lambda ^2\right]
\end{equation}
We can substitute equations (21) and (23) into equation (16) and eliminate the $p_\lambda$ denominator factors to arrive at a polynomial equation in $\lambda$ and $\{ x_k\}$, which we term the \emph{exponential Hessian determinant}.

The equations (10), (16), and (19) make up a system of $n+2$ multivariate polynomial equations in $2n+1$ variables; these are $\lambda$, $\{ x_k\}$, and $\{ X_k\}$. Using a Gr{\"o}bner basis calculation \cite{cox06}, we can reduce this system to a single equation of $n$ spatial variables $\{ X_k\}$. Unfortunately the exponential Hessian determinant is often prohibitively large and the system requires significant computational resources to solve. However, we present a more feasible na{\"i}ve method of bifurcation curve computation which, while not rigorously supported, generates curves that bear striking visual resemblance to the quantum walk boundaries. Consider the polynomial system $f(x)=\sum _{k=0}^n a_kx^k=0$ and $g(x)=\sum _{k=0}^n b_kx^k=0$. We wish to find a resultant multivariate polynomial $F(a_k,b_k)$ such that selections of coefficients $\{ a_k,b_k\}$ which admit simultaneous solutions of the polynomial system in $x$ also satisfy the equation $F(a_k,b_k)=0$. Such a resultant may be computed using the determinant of a $2n\times 2n$ Sylvester matrix \cite{sylvester12}: 
$$\text{Res}(f,g;x)=F(a_k,b_k)=\det\begin{bmatrix} a_n & a_{n-1} & \hdots & a_0 & 0 & ~ & ~ & ~ \\ b_n & b_{n-1} & \hdots & b_0 & 0 & \ddots & ~ & ~ \\ 0 & a_n & \hdots & a_1 & a_0 & \ddots & ~ & ~ \\ 0 & b_n & \hdots & b_1 & b_0 & \ddots & ~ & ~ \\ ~ & ~ & \ddots & \ddots & \ddots & \ddots & ~ & ~ \\ ~ & ~ & ~ & ~ & ~ & a_n & \hdots & a_0 \\ ~ & ~ & ~ & ~ & ~ & b_n & \hdots & b_0\end{bmatrix}$$
In this notation, $x$ is the variable being cancelled. There are two shortcomings with this formula. First, this resultant will often overrepresent solutions in the system in the sense that solutions of $F(a_k,b_k)=0$ in $\{ a_k,b_k\}$ may not admit solutions in the corresponding polynomial system. For example, consider the system $f(x)=ax+b=0$ and $g(x)=cx+d=0$ such that the resultant satisfies $F(a,b,c,d)=ad-bc=0$. If we let $a=c=0$, then the resultant equation is trivially satisfied, but any nonzero choice of $b$ or $d$ leads to a polynomial system with no solutions. The second shortcoming of this procedure is that the resultant will not take into account any \emph{a priori} restrictions on the cancelled variable. For instance in our current quantum walk example, we will require that the cancelled variables satisfy $|\lambda |=|x_k|=1$. It is often difficult to discern which portions of the generated bifurcation curves satisfy these conditions, as these variables are absent from the resultant. Though this method overrepresents solutions of the polynomial system, it will not miss any of the solutions and we use our non-rigorous judgment to hypothesize which ones truly exist in the system.

We use a simple extension of this method to solve a larger system of polynomial equations. Suppose we have a system of $n$ polynomial equations in $n$ variables with a set of variable coefficients. Let us write these polynomials as $p_{0,k}(x_1,...,x_n)$ where $1\le k\le n$. Using the Sylvester matrix determinant, let $p_{1,k}(x_1,...,x_{n-1})=\text{Res}(p_{0,k},p_{0,n};x_n)$ for $1\le k\le n-1$, in other words we choose a \emph{base polynomial} $p_{0,n}$ and compute resultants with the remaining polynomials in the system by cancelling $x_n$. The new system $\{ p_{1,k}\}$ has $n-1$ equations and $n-1$ variables. We can continue inductively to arrive at a total multivariate resultant polynomial in the coefficient variables. As with the Gr{\"o}bner basis calculation, even small systems in multiple variables can lead to extremely large resultant polynomials. To combat this, we factor the intermediate polynomials in the system to create a tree of possible solutions, these putative solutions being more feasible to derive. Also, depending on the structure of the system, different choices of base polynomials as well as different orders of variable cancellations can often lead to significant changes in computation time. 

Up to this point, the na{\"i}ve method we have described for computing bifurcation curves is legitimate, we need only take care to ensure that we judiciously select correct bifurcation curves from the overrepresentation provided. However, the exponential Hessian determinant still provides a massive roadblock and renders this na{\"i}ve method just as intractible as the Gr{\"o}bner basis method. It has been observed that replacing the exponential Hessian determinant with the far simpler equation $p_\lambda =0$ results in bifurcation curve solutions which visually bound the regions of polynomial decay, though it is not clear how these curves can be rigorously justified. If we let $p_\lambda =0$, then there are no restrictions on the spatial variables $X_j$ in equation (20), and these are the variables of interest. A Gr{\"o}bner basis calculation would fail to render spatial bifurcation curves in this case while the na{\"i}ve method generates outputs. We have found that letting $p_\lambda=0$ be the first base polynomial, and cancelling the variables $\{ x_k\}$ before cancelling $\lambda$ leads to more digestible factors for the algorithm. In the subsequent section, we will clearly state when the displayed bifurcation curves result from the Gr{\"o}bner basis calculation and when they are derived from the na{\"i}ve method.

\section{Examples} \label{sec:4}
In this section, we compute parameterizations of regions of polynomial decay for five different two-dimensional quantum walks and compute bifurcation curves where possible. In the following cases, characteristic polynomials of the multiplier matrix will often be symmetric quartic polynomials. Suppose we have a characteristic polynomial
$$p_0(\lambda ;x,y)=\lambda ^4+x(\theta )\lambda ^3+y(\theta )\lambda ^2+x(\theta )\lambda +1$$
such that $x$ and $y$ are functions of a vector $\theta$. This polynomial can be factored as:
$$p_0(\lambda ;x,y)=(\lambda ^2+a(\theta )\lambda +1)(\lambda ^2+b(\theta )\lambda +1)$$
where $a+b=x$ and $ab+2=y$, otherwise $a(\theta)=\frac{1}{2}\left[ x(\theta )\pm\sqrt{x(\theta )^2-4y(\theta )+8}\right]$ and $b(\theta )$ takes the opposite sign. These quadratic factors allow for an explicit representation of the eigenvalues, but recall that $\lambda =e^{iH}$ and we are searching for $\nabla H$. Using implicit differentiation and previous equations, we find that the following holds:
\begin{equation}
\nabla H(\theta)=\frac{a(\theta )\nabla x(\theta )-\nabla y(\theta )}{(2a(\theta )-x(\theta ))\sqrt{4-a(\theta )^2}}
\end{equation}
This formula will grant us a parametric representation for the region of polynomial decay in these examples.

\subsection{Four-State Grover Walk}
We first consider the two-dimensional four-state Grover walk $(\Z ^2,C_2,G_4)$. Recall that the $4\times 4$ Grover matrix is written as:
$$G_4=\frac{1}{2}\begin{bmatrix} -1 & 1 & 1 & 1 \\ 1 & -1 & 1 & 1 \\ 1 & 1 & -1 & 1 \\ 1 & 1 & 1 & -1\end{bmatrix}$$
We write the corresponding multiplier matrix $M(\theta _1,\theta _2)$ as:
$$M(\theta _1,\theta _2)=\frac{1}{2}\begin{bmatrix} -e^{i\theta _1} & e^{i\theta _1} & e^{i\theta _1} & e^{i\theta _1} \\ e^{-i\theta _1} & -e^{-i\theta _1} & e^{-i\theta _1} & e^{-i\theta _1} \\ e^{i\theta _2} & e^{i\theta _2} & -e^{i\theta _2} & e^{i\theta _2} \\ e^{-i\theta _2} & e^{-i\theta _2} & e^{-i\theta _2} & -e^{-i\theta _2}\end{bmatrix}$$
The characteristic polynomial of the multiplier matrix thus satisfies:
$$p_0(\lambda ;\theta _1,\theta _2)=(\lambda ^2-1)(\lambda ^2+(\cos\theta _1+\cos\theta _2)\lambda +1)$$
We pause to note that the solution $\lambda =\pm 1$ causes a breakdown in the stationary phase approximation in that the only critical point that exists corresponding to this eigenvalue is at $(X_1,X_2)=(0,0)$ and is degenerate. This phenomenon is known as localization and has been explored by several authors \cite{inui04} \cite{inui05} \cite{konno10} \cite{stefanak12}; we will not elaborate any further on it in this paper. We will term constant solutions to the characteristic polynomial as \emph{trivial}.

As the non-trivial portion of the characteristic polynomial is quadratic, we can explicitly solve for the non-trivial eigenvalues and use a reduced version of equation (24) to construct a parameterization of the region of polynomial decay:
\begin{equation}
(X_1,X_2)=\left(\frac{\sin\theta _1}{\sqrt{4-(\cos\theta _1+\cos\theta _2)^2}},\frac{\sin\theta _2}{\sqrt{4-(\cos\theta _1+\cos\theta _2)^2}}\right)
\end{equation}
It was shown in Kuklinski \cite{kuklinski17} that this parameterization traces the circle $2X_1^2+2X_2^2\le 1$ in $\R ^2$, albeit in an atypical way.

In this example, we can in fact solve for the bifurcation curve. By letting $x_1=e^{i\theta _1}$ and $x_2=e^{i\theta _2}$, we can write the characteristic equation in a different way:
$$p(\lambda ;x_1,x_2)=2x_1x_2\lambda ^2+(x_1x_2+1)(x_1+x_2)\lambda +2x_1x_2=0$$
The corresponding exponential Hessian determinant is small enough that the system may be efficiently reduced via Gr{\"o}bner basis computation. The result is as we expect:
\begin{equation}
2X_1^2+2X_2^2=1
\end{equation}

\begin{figure}
\begin{center}
\includegraphics[width=4cm]{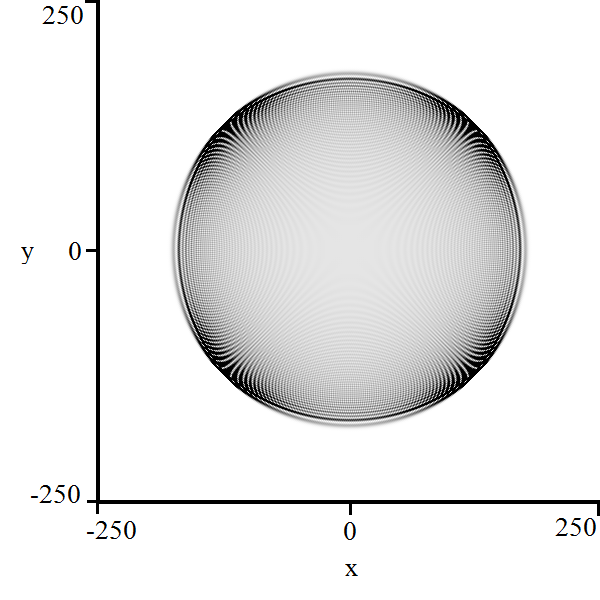}
\hspace{1cm}
\includegraphics[width=4cm]{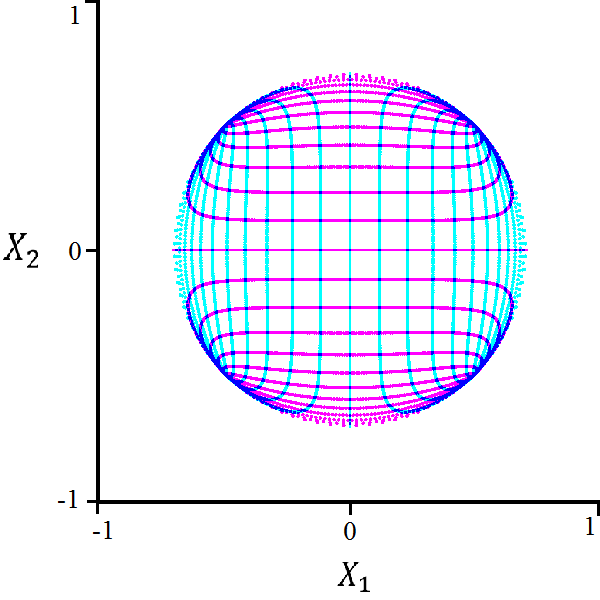}
\hspace{1cm}
\includegraphics[width=4cm]{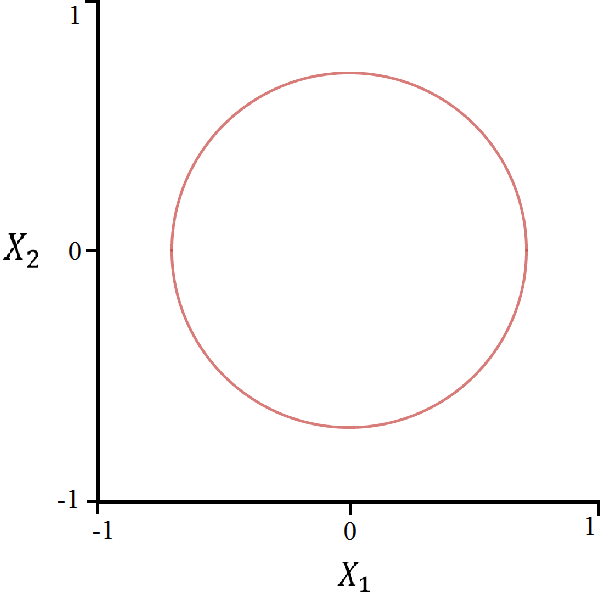}
\end{center}
\caption{(\emph{Left}) 250 steps of the four-state Grover walk in initial position $\frac{1}{2}|0,0\rangle \left(|R\rangle +|L\rangle -|U\rangle -|D\rangle\right)$ (\emph{Center}) Parameterization of the four-state Grover walk region of polynomial decay (\emph{Right}) Bifurcation curve of the four-state Grover walk.}
\end{figure}

\subsection{Five-State Grover Walk}
We explore a variant of the four-state Grover walk \cite{ampadu112} with the operator $Q\leftrightarrow (\Z ^2,\tilde{C_2},G_5)$ The $5\times 5$ Grover matrix is written as:
$$G_5=\frac{1}{5}\begin{bmatrix} -3 & 2 & 2 & 2 & 2 \\ 2 & -3 & 2 & 2 & 2 \\ 2 & 2 & -3 & 2 & 2 \\ 2 & 2 & 2 & -3 & 2 \\ 2 & 2 & 2 & 2 & -3\end{bmatrix}$$
We write the corresponding multiplier matrix $M(\theta _1,\theta _2)$ as:
$$M(\theta _1,\theta _2)=\frac{1}{5}\begin{bmatrix} -3e^{i\theta _1} & 2e^{i\theta _1} & 2e^{i\theta _1} & 2e^{i\theta _1} & 2e^{i\theta _1} \\ 2e^{-i\theta _1} & -3e^{-i\theta _1} & 2e^{-i\theta _1} & 2e^{-i\theta _1} & 2e^{-i\theta _1} \\ 2 & 2 & -3 & 2 & 2 \\ 2e^{i\theta _2} & 2e^{i\theta _2} & 2e^{i\theta _2} & -3e^{i\theta _2} & 2e^{i\theta _2} \\ 2e^{-i\theta _2} & 2e^{-i\theta _2} & 2e^{-i\theta _2} & 2e^{-i\theta _2} & -3e^{-i\theta _2}\end{bmatrix}$$
The characteristic polynomial of this multiplier matrix satisfies
\begin{align*}
    & p_0(\lambda ;\theta _1,\theta _2) \\ 
    & =(\lambda -1)\left( \lambda ^4+\frac{2}{5}(3c_1+3c_2+4)\lambda ^3+\frac{2}{5}(4c_1+4c_2+2c_1c_2+5)\lambda ^2+\frac{2}{5}(3c_1+3c_2+4)\lambda +1\right) =0
\end{align*}
where $c_1=\cos\theta _1$ and $c_2=\cos\theta _2$. Notice that the eigenvalue $\lambda =1$ leads to localization in this walk as well.

\begin{figure}
\begin{center}
\includegraphics[width=5.3cm]{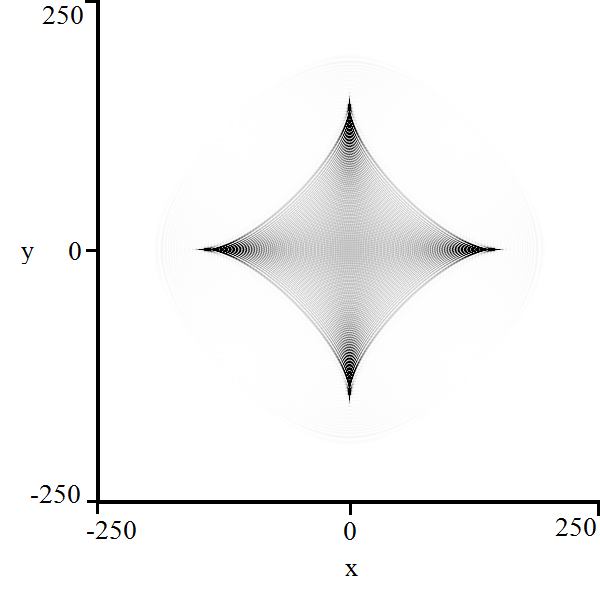}
\hspace{1cm}
\includegraphics[width=5.3cm]{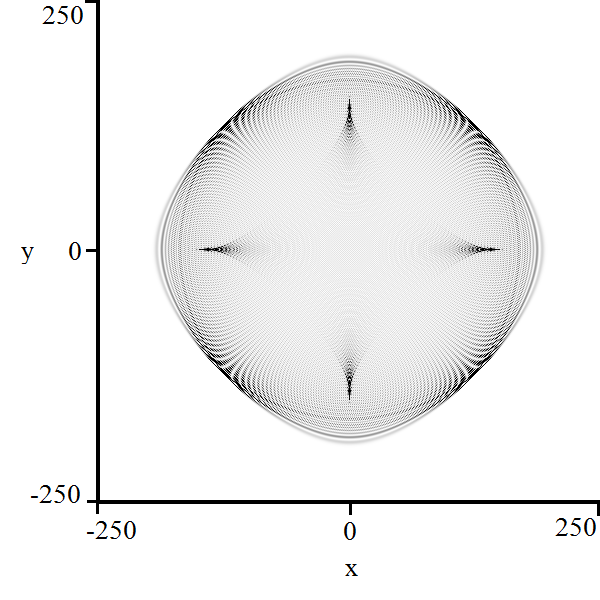}
\end{center}
\caption{(\emph{Left}) 250 steps of the five-state Grover walk in initial position $\frac{1}{2\sqrt{5}}|0,0\rangle \left(|R\rangle +|L\rangle +|U\rangle +|D\rangle -|S\rangle\right)$ (\emph{Right}) 250 steps of the five-state Grover walk in initial position $\frac{1}{2}|0,0\rangle \left(|R\rangle +|L\rangle -|U\rangle -|D\rangle\right)$.}
\end{figure}

Using equation (24), we can write an explicit parameterization of the region of polynomial decay:
\begin{align}
(X_1,X_2)= & \left( \frac{3as_1-4s_1-2s_1c_2}{\sqrt{(9c_1^2+9c_2^2-2c_1c_2-16c_1-16c_2+16)(4-a^2)}}, \right. \notag \\ & \left. \qquad \qquad \qquad \qquad \qquad  \frac{3as_2-4s_2-2s_2c_1}{\sqrt{(9c_1^2+9c_2^2-2c_1c_2-16c_1-16c_2+16)(4-a^2)}} \right)
\end{align}
Here, $s_1=\sin\theta_1$, $s_2=\sin\theta _2$, and $a=\frac{1}{5}\left( 3c_1+3c_2+4\pm\sqrt{9c_1^2+9c_2^2-2c_1c_2-16c_1-16c_2+16}\right)$. Notice that in this case, the choice of plus/minus in $a(\theta _1,\theta _2)$ results in different regions of polynomial decay.

To find the bifurcation curves of this quantum walk, we rewrite the characteristic polynomial:
\begin{align*}
p(\lambda ;x_1,x_2) &= 5x_1x_2\lambda ^4+(3x_1^2x_2+3x_1x_2^2+8x_1x_2+3x_1+3x_2)\lambda ^3 \\
	&+ (x_1^2x_2^2+4x_1^2x_2+x_1^2+4x_1x_2^2+10x_1x_2+4x_1+x_2^2+4x_2+x_2^2+1)\lambda ^2 \\
	&+(3x_1^2x_2+3x_1x_2^2+8x_1x_2+3x_1+3x_2)\lambda +5x_1x_2=0
\end{align*}
In this example, the exponential Hessian determinant is several pages long, so a Gr{\"o}bner basis calculation is computationally infeasable. We choose to illustrate the na{\"i}ve method for this system. Let $P_1=p$, $P_2=p_1x_1+\lambda p_\lambda X_1$, $P_3=p_2x_2+\lambda p_\lambda X_2$, and $P_4=p_\lambda$. We first cancel $x_1$ from this system by letting $P_4$ be the base polynomial:
$$\text{Res}(P_1,P_4,x_1)=(\lambda -1)^2(\lambda +1)^2Q_1$$
$$\text{Res}(P_2,P_4,x_1)=\lambda^2(\lambda +1)^2Q_2^2$$
$$\text{Res}(P_2,P_4,x_1)=\lambda^2(\lambda +1)^2Q_3^2$$
The $Q_k$ polynomials are irreducible. We can omit factors of $\lambda$ in the second two equations as $|\lambda |=1$. If we substitute the factor $\lambda =\pm 1$ from one of these equations, substitute into the remaining two, and then cancel $x_2$ from the resulting system, the generated curves do not visually fit the system, so we ignore this option and proceed with solving the system $\{ Q_1,Q_2,Q_3\}$. By choosing $Q_2$ to be the base polynomial and eliminating $x_2$ from the system, we find:
$$\text{Res}(Q_1,Q_2,x_2)=\lambda ^{40}X_2^{12}(\lambda -1)^{14}(\lambda +1)^{32}R_1$$
$$\text{Res}(Q_2,Q_3,x_1)=\lambda ^{40}(15\lambda ^4+20\lambda ^3+58\lambda ^2+20\lambda +15)^2(\lambda +1)^{32}$$
If we substitute $\lambda =-1$ from the second equation into $R_1$, then we find:
\begin{align}
0 &= 405(X_1^8+X_2^8)-648(X_1^6+X_2^6)+(378-180X_1^2X_2^2)(X_1^4+X_2^4) \\
	&+(1416X_1^2X_2^2-96)(X_1^2+X_2^2)+830X_1^4X_2^4-596X_1^2X_2^2+9\nonumber
\end{align}
The remaining factors from the second polynomial do not satisfy $|\lambda |=1$ so we ignore these. Although this is not proof, equation (28) visually replicates the boundary of both regions of polynomial decay. Notice that although we have found two distinct regions of polynomial decay in the parametric representation, this irreducible bifurcation curve traces the boundaries of both regions. We note that the outer curve has a maximum distance of $\sqrt{\frac{3}{5}}$ and a minimum distance of $\frac{1}{\sqrt{2}}$ from the origin, while the inner curve has a maximum distance of $\frac{1}{\sqrt{3}}$ and a minimum distance of $\frac{1}{\sqrt{10}}$ from the origin. These maxima are attained on the cardinal axes, and the minima are attained on the axes $y=\pm x$.

\begin{figure}
\begin{center}
\includegraphics[width=4cm]{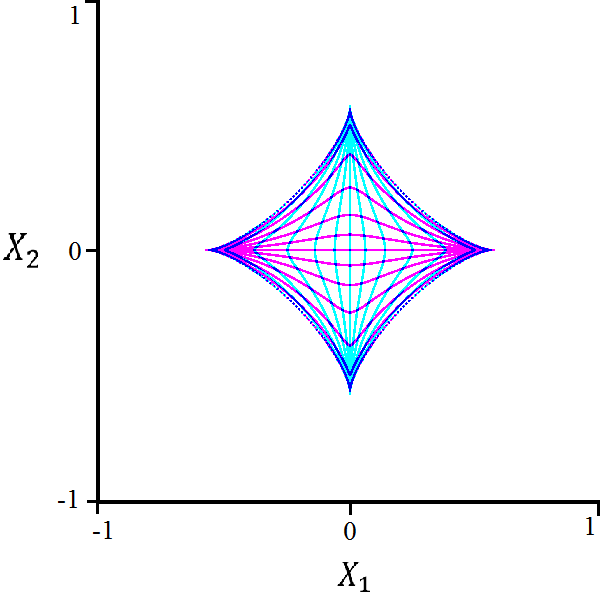}
\hspace{1cm}
\includegraphics[width=4cm]{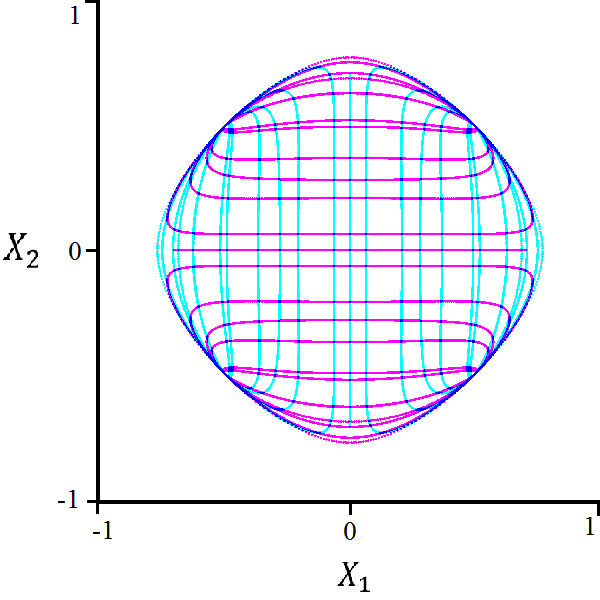}
\hspace{1cm}
\includegraphics[width=4cm]{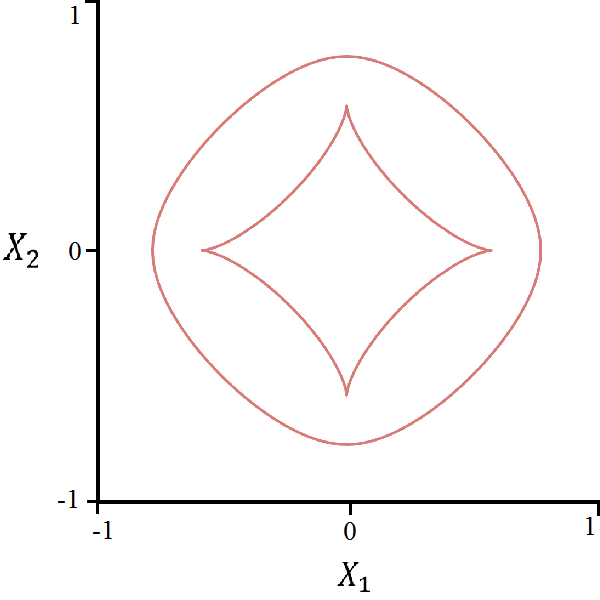}
\end{center}
\caption{(\emph{Left/Center}) Parameterizations of the five-state Grover walk regions of polynomial decay (\emph{Right}) Bifurcation curve prediction of the five-state Grover walk.}
\end{figure}

\subsection{Triangular Quantum Walk}
We now analyze the triangular Grover walk with operator $Q\leftrightarrow (\Z ^2,\Sigma ,G_3)$ where \\ 
$\Sigma =\{ (0,1),(-1,-1),(1,-1)\}$. The multiplier matrix of this operator may be written as:
$$M(\theta _1,\theta _2)=\frac{1}{3}\begin{bmatrix} -e^{i\theta _2} & 2e^{i\theta _2} & 2e^{i\theta _2} \\ 2e^{-i(\theta _1+\theta _2)} & -e^{-i(\theta _1+\theta _2)} & 2e^{-i(\theta _1+\theta _2)} \\ 2e^{i(\theta _1-\theta _2)} & 2e^{i(\theta _1-\theta _2)} & -e^{i(\theta _1-\theta _2)}\end{bmatrix}$$
Letting $x_k=e^{i\theta _k}$, the eigenvalues of this matrix satisfy the following equation:
$$p(\lambda ;x_1,x_2)=3x_1x_2^2\lambda ^3+x_2(x_1^2+x_1x_2^2+1)\lambda ^2-(x_1^2x_2^2+x_1+x_2^2)\lambda -3x_1x_2=0$$
The structure of this polynomial does not lend itself to an analytic solution of $\lambda (\theta )$ without invoking the cubic formula \cite{guilbeau30}. We do not write the formula here, but we may graphically display this parameterization.

Furthermore, the Hessian determinant is too large to facilitate a Gr{\"o}bner basis calculation for the bifurcation curves. However, using the na{\"i}ve elimination procedure outputs the bifurcation curve:
\begin{equation}
4X_1^2+3X_2^2+2X_2-1=0
\end{equation}
This equation represents an ellipse centered at $(X_1,X_2)=(0,-\frac{1}{3})$ with vertical major axis length $\frac{4}{3}$ and horizontal minor axis length $\frac{2}{\sqrt{3}}$.

\begin{figure}
\begin{center}
\includegraphics[width=4cm]{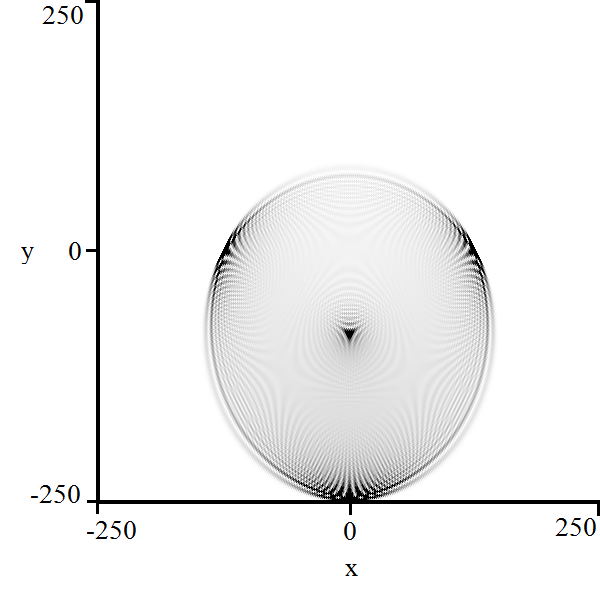}
\hspace{1cm}
\includegraphics[width=4cm]{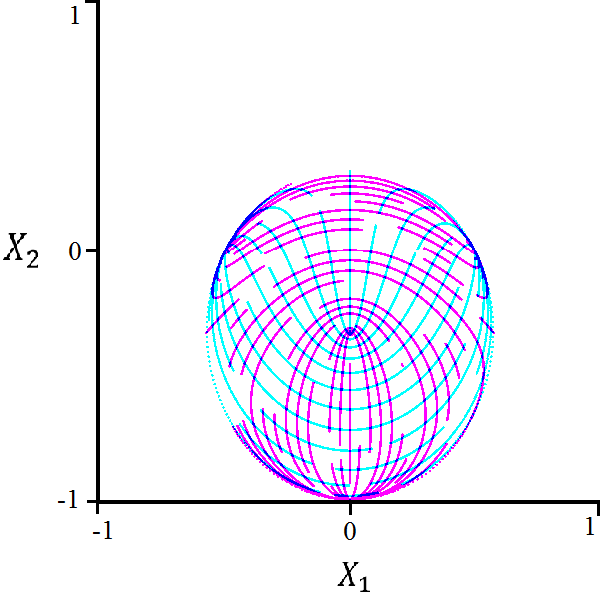}
\hspace{1cm}
\includegraphics[width=4cm]{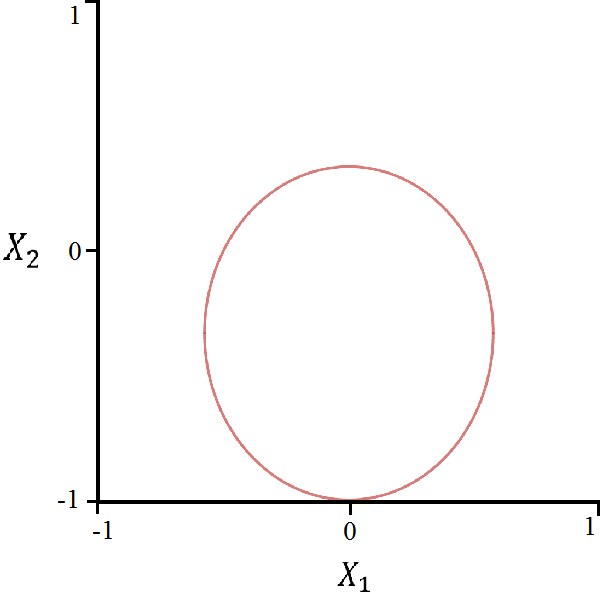}
\end{center}
\caption{(\emph{Left}) 250 steps of the triangular quantum walk in initial position $\frac{1}{\sqrt{3}}|0,0\rangle \left(i|U\rangle +|RD\rangle -|LD\rangle\right)$ (\emph{Center}) Parameterization of the triangular quantum walk region of polynomial decay (\emph{Right}) Bifurcation curve prediction of the triangular quantum walk.}
\end{figure}

\subsection{Hexagonal Quantum Walk}
We now illustrate an example of a quantum walk which traverses a hexagonal lattice on $\Z ^2$. The hexagonal or honeycomb lattice has been the subject of a few quantum walk investigations \cite{jafarizadeh07} \cite{lyu15}. Let us define the set:
$$\Sigma =\{ (2,0),(-1,1),(-1,-1),(-2,0),(1,-1),(1,1)\}$$
Consider the quantum walk operator $Q\leftrightarrow (\Z ^2,\Sigma ,X\otimes G_3)$ where $X$ is the Pauli-$X$ gate \cite{griffiths82} $X=\begin{bmatrix} 0 & 1 \\ 1 & 0\end{bmatrix}$. In this case, amplitudes will travel on two separate hexagonal lattices. To be clear, this hexagonal Grover walk takes place on a subset of $\Z ^2$ spanned by the elements of $\Sigma$. The multiplier matrix of this quantum walk takes the form:
$$M(\theta _1,\theta _2)=\frac{1}{3}\begin{bmatrix}0 & 0 & 0 & -e^{2i\theta _1} & 2e^{2i\theta _1} & 2e^{2i\theta _1} \\ 0 & 0 & 0 & 2e^{i(\theta _2-\theta _1)} & -e^{i(\theta _2-\theta _1)} & 2e^{i(\theta _2-\theta _1)} \\ 0 & 0 & 0 & 2e^{-i(\theta _1+\theta _2)} & 2e^{-i(\theta _1+\theta _2)} & -e^{-i(\theta _1+\theta _2)} \\ -e^{-2i\theta _1} & 2e^{-2i\theta _1} & 2e^{-2i\theta _1} & 0 & 0 & 0 \\ 2e^{i(\theta _1-\theta _2)} & -e^{i(\theta _1-\theta _2)} & 2e^{i(\theta _1-\theta _2)} & 0 & 0 & 0 \\ 2e^{i(\theta _1+\theta _2)} & 2e^{i(\theta _1+\theta _2)} & -e^{i(\theta _1+\theta _2)} & 0 & 0 & 0\end{bmatrix}$$

The characteristic polynomial of the multiplier matrix of this quantum walk operator is written as:
$$p_0(\lambda ,\theta _1,\theta _2)=(\lambda ^2-1)\left( \lambda ^4-\frac{2}{9}\left[ 4\cos{2\theta _2}+8\cos{3\theta _1}\cos{\theta _2}-3\right]\lambda ^2+1\right)$$
The factor $\lambda ^2-1$ indicates that localization is present in this walk. The remaining eigenvalues satisfy the following equation:
$$p(\lambda ;x_1,x_2)=9x_1^3x_2^2\lambda ^4-(4x_1^6x_2^3+4x_1^6x_2+4x_1^3x_2^4-6x_1^3x_2^2+4x_1^3+4x_2^3+4x_2)\lambda ^2+9x_1^3x_2^2=0$$
Since the characteristic polynomial is quadratic in $\lambda ^2$, we can efficiently parameterize the region of polynomial decay:
\begin{equation}
(X_1,X_2)=\left(\frac{12\sin{3\theta _1}\cos\theta _2}{\sqrt{81-(4\cos{2\theta _2}+8\cos{3\theta _1}\cos\theta _2-3)^2}},\frac{4\sin{2\theta _2}+4\cos{3\theta _1}\sin\theta _2}{\sqrt{81-(4\cos{2\theta _2}+8\cos{3\theta _1}\cos\theta _2-3)^2}}\right)
\end{equation}

Again, the characteristic polynomial is too large for a Gr{\"o}bner basis calculation, but the na{\"i}ve method leads to two possible bifurcation curves:
\begin{align}
X_1^2+3X_2^2-1 &= 0 \\
X_1^2+3X_2^2-2 &= 0
\end{align}
These equations represent two concentric ellipses with major axis of length $\sqrt{3}$ times the length of the minor axis. It seems likely that the larger ellipse in equation (31) is a bifurcation curve of the system, but it is uncertain whether the smaller ellipse in equation (32) is a legitimate bifurcation curve.

\begin{figure}
\begin{center}
\includegraphics[width=4cm]{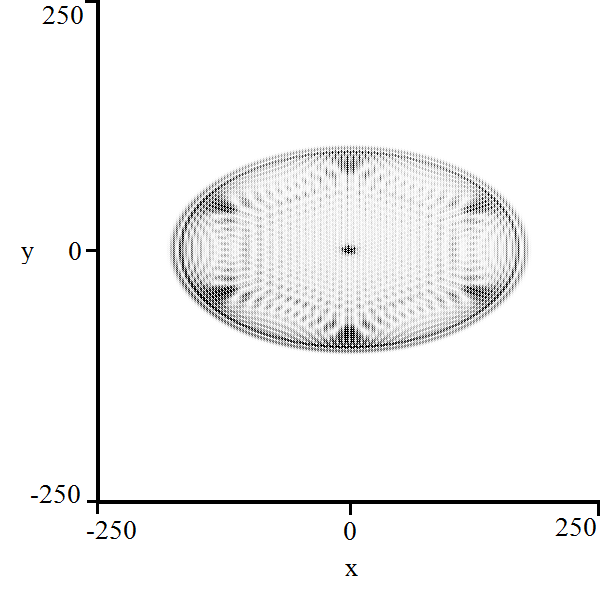}
\hspace{1cm}
\includegraphics[width=4cm]{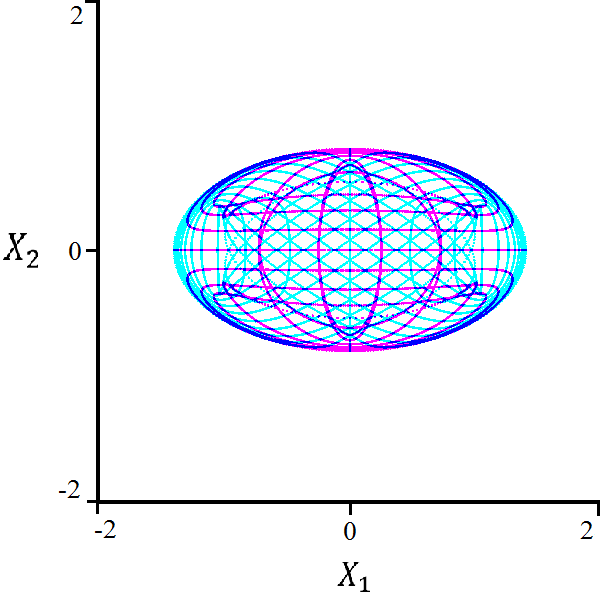}
\hspace{1cm}
\includegraphics[width=4cm]{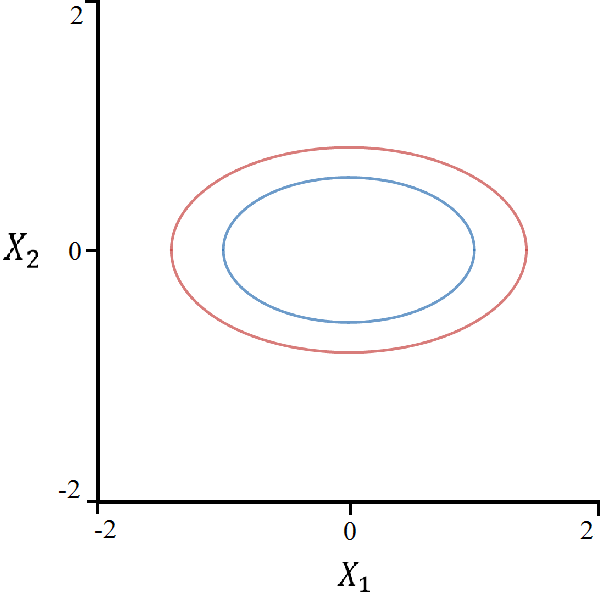}
\end{center}
\caption{(\emph{Left}) 125 steps of the hexagonal quantum walk in initial position $\frac{1}{\sqrt{6}}|0,0\rangle \left(|R\rangle +|LU\rangle +|LD\rangle +|L\rangle +|RD\rangle +|RU\rangle\right)$ (\emph{Center}) Parameterization of the hexagonal quantum walk region of polynomial decay (\emph{Right}) Bifurcation curve prediction of the hexagonal quantum walk.}
\end{figure}

\subsection{Four-State Hadamard Walk}
In the final example we consider a quantum walk governed by a different unitary matrix. Let $Q\leftrightarrow (\Z ^2,C_2,H\otimes H)$ such that the corresponding multiplier matrix becomes:
$$M(\theta _1,\theta _2)=\frac{1}{2}\begin{bmatrix} e^{i\theta _1} & e^{i\theta _1} & e^{i\theta _1} & e^{i\theta _1} \\ e^{-i\theta _1} & -e^{-i\theta _1} & e^{-i\theta _1} & -e^{-i\theta _1} \\ e^{i\theta _2} & e^{i\theta _2} & -e^{i\theta _2} & -e^{i\theta _2} \\ e^{-i\theta _2} & -e^{-i\theta _2} & -e^{-i\theta _2} & e^{-i\theta _2}\end{bmatrix}$$
The characteristic polynomial of this matrix becomes:
$$p_0(\lambda ,\theta _1,\theta _2)=\lambda ^4-i(\sin\theta _1+\sin\theta _2)\lambda ^3-(\cos (\theta _1+\theta _2)+1)\lambda ^2+i(\sin\theta _1+\sin\theta _2)\lambda +1=0$$
This is not a symmetric quartic polynomial, but the parametrization of the region of polynomial decay may stille be solved using a modification of equation (24):
\begin{equation}
(X_1,X_2)=\left(\frac{as_1+s_{1+2}}{\sqrt{((c_1+c_2)^2-4c_{1+2}+4)(4-a^2)}},-\frac{as_2+s_{1+2}}{\sqrt{((c_1+c_2)^2-4c_{1+2}+4)(4-a^2)}}\right)
\end{equation}
Here, we let $s_{1+2}=\sin (\theta _1+\theta _2)$, $c_{1+2}=\cos (\theta _1+\theta _2)$, and $a=-\frac{1}{2}\left[ c_1+c_2\pm\sqrt{(c_1+c_2)^2-4c_{1+2}+4}\right]$.

By letting $x_1=e^{i\theta _1}$ and $e^{i\theta _2}$, we can rewrite the characteristic polynomial:
$$p(\lambda ;x_1,x_2)=2x_1x_2\lambda ^4-(x_1x_2+1)(x_1-x_2)\lambda ^3-(x_1x_2+1)^2\lambda ^2+(x_1x_2+1)(x_1-x_2)\lambda +2x_1x_2=0$$
This characteristic polynomial is too large for a Gr{\"o}bner basis calculation, and even the na{\"i}ve method cannot generate a complete set of outputs. However, this algorithm is capable of generating the equations for the two main ellipses in the region of polynomial decay:
\begin{align}
3X_1^2-2X_1X_2+2X_1+3X_2^2-2X_2 &= 0 \\
3X_1^2-2X_1X_2-2X_1+3X_2^2+2X_2 &= 0
\end{align}
The major axes of these ellipses are parallel to the line $x=y$ and have length 1 while the minor axes have length $\frac{\sqrt{2}}{2}$. Though it did not appear in the calculation, we also predict that the bifurcation curve set also includes a rhombus and a $16^\text{th}$ order algebraic curve. 

\begin{figure}
\begin{center}
\includegraphics[width=4cm]{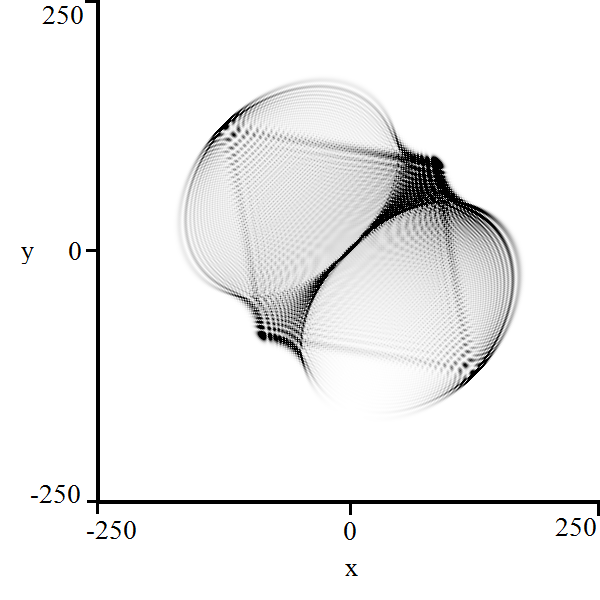}
\hspace{1cm}
\includegraphics[width=4cm]{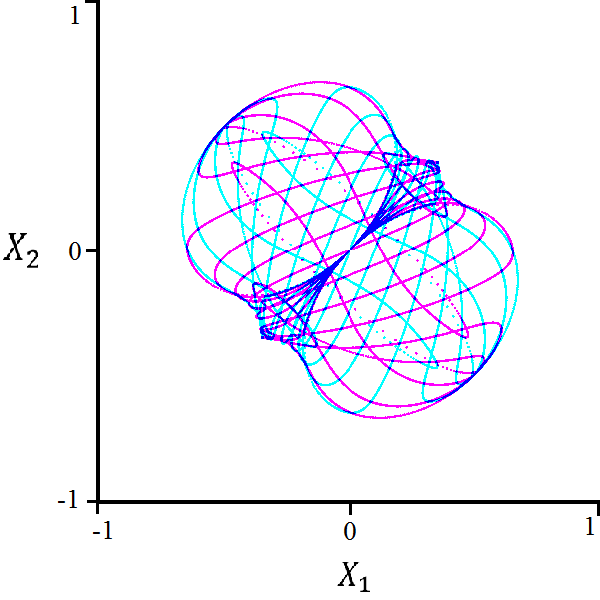}
\hspace{1cm}
\includegraphics[width=4cm]{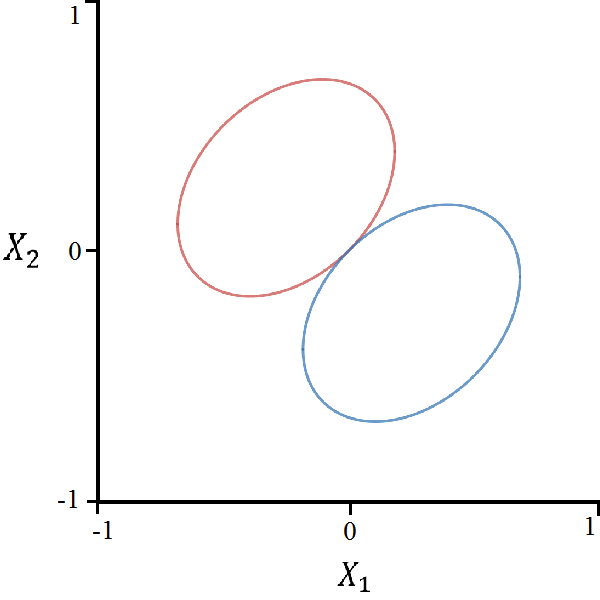}
\end{center}
\caption{(\emph{Left}) 250 steps of the Hadamard walk in initial position $\frac{1}{\sqrt{2}}|0,0\rangle \left(|R\rangle +|U\rangle\right)$ (\emph{Center}) Parameterization of the Hadamard walk region of polynomial decay (\emph{Right}) Partial bifurcation curve prediction of the Hadamard walk.}
\end{figure}

\section{Conclusion}
In this paper we have detailed a process to compute bifurcation curves of two-dimensional quantum walks, as well as describe a non-rigorous algorithm to trace bifurcation curves for more complicated examples which are difficult to solve analytically. In addition, we have provided parameterizations of the regions of polynomial decay. These methods are not unique to two-dimensional quantum walks, and with sufficient computational resources could potentially be extended to computing bifurcation surfaces for higher-dimensional quantum walks. 

\nocite{*}
\bibliographystyle{eptcs}
\bibliography{biblio}
\end{document}